# FORC and Micromagnetism Approach to the Domain Structure of Cobalt Antidot Arrays

S. Michea[1], J. Briones[2,4], J. L. Palma[2,4], R. Lavín[3], J. Escrig[2,4], R. Rodríguez-Suárez[1] and J. C. Denardin[2,4]

[1] Facultad de Física, Pontificia Universidad Católica de Chile, Casilla 306, Santiago, Chile.
[2] Departamento de Física, Universidad de Santiago de Chile, Av. Ecuador 3493, Santiago, Chile.
[3] Facultad de Ingeniería, Universidad Diego Portales, Ejército 441, Santiago, Chile.
[4] Centro para el Desarrollo de la Nanociencia y la Nanotecnología, Av. Ecuador 3493, Santiago, Chile.

E-mail: smichea@uc.cl

**Abstract.**
We study the influence of the porosity on the domain structure of cobalt antidots thin films with controlled and circular defects of 20, 40 and 60 nm of diameter. Micromagnetic simulations, combined with First-order reversal curves analysis of classical magnetometry measurements, have been used to track the evolution of the magnetic domain configurations. The found coercivity enhancement with the increase of the pore diameter is correlated to the domain reversibility. Moreover, we found that when the pores diameter increases the domain-domain interactions become dominant.

## 1. Introduction

Nanostructured magnetic elements have received much attention from the scientific community in the last two decades due to their potential applications, ranging from sensors for the electronic and electromechanical industry to the storage media for the magnetic recording industry [1, 2, 3]. Such magnetic nanostructures are possible due to the recent advances in both lithographic techniques [4] (top-down approach) and self-organization processes [5, 6, 7, 8] (bottom-up approach). These technologies enable scientists and engineers to control the size and geometry of the magnetic materials obtaining thus a broad spectrum of magnetic properties. Among the magnetic nanostructures that can be obtained by using self-organized processes there are the so-called magnetic antidots (i. e. holes arrangement embedded into a continuous magnetic film). In effect, the use of self-organized ordered nanoporous alumina membranes as templates has allowed the fabrication of magnetic antidots films with high hexagonal order and controlled pores distribution [9]. Magnetic antidots are a topic of intense current research since they are promising candidates for high-density information storage as well as a very exciting topic in fundamental physics [10, 11, 12, 13, 14, 15, 16, 17].

In an antidot array, magnetic features such as coercive field, anisotropy axes and reversal mechanisms, among others, can be tailored by tuning the geometric parameters of the array [18, 19, 20]. Such parameters are: the shape and diameter of the hole, the distance between holes and their spatial distribution. Besides, in the absence of a regular arrangement of the holes, the porosity can be used as a geometric parameter. In this work, we propose in an experimental way, a schematic study about the influence of the porosity on the magnetic properties of Cobalt antidots. Micromagnetic simulations performed in these systems enabled us track to the reversal modes and the magnetization state of the hysteresis loop in the entire film. Complementary, first order reversal curves (FORCs) performed on the antidots films add information about the inner interactions that may arise into these systems.

## 2. Methodology

### 2.1. *Fabrication of cobalt antidots*

Nanoporous alumina membranes were prepared from a 0.32 mm thickness aluminum foil (Good-Fellow, 99.999%) by using the so-called two-step anodization technique [7]. Prior to the anodization processes the aluminum foils were cleaned with acetone, isopropanol and distilled water, and then electropolished for 5-10 minutes in an ethanol:perchloric acid (3:1) solution under a 25 V of voltage at 4°C. After this treatment the samples were submitted to a first anodization, at 40 V and 45 V for 8 hours in an oxalic acid solution at 20°C, in order to obtain large pore sizes, and in a 0.3 M sulphuric acid solution under a 20 V voltage at 20°C, in the case of smaller pore diameters. After the first anodization step, the anodized layer was etched with a solution at room temperature during 12 hours. This solution is composed of 7 gr of $H_3PO_4$, 1.8 gr of $H_3CrO_4$ and adding $H_2O$ up to complete 1000 ml. The ordered pore arrangement was achieved with a 6 hours long second anodization step performed under same conditions than the first one. The pore diameters for each sample obtained were 20, 40, and 60 nm. The inter-pore distance is 50 nm in the case of samples made by sulphuric acid and 100 nm for oxalic acid. In all cases, the membranes show a hexagonal order and good uniformity in both, the pore diameter and the distance between pores. The cobalt antidots were fabricated by magnetron sputtering deposition of a 28 nm cobalt layer on the top of the membranes with three different pore diameters, replicating the well-ordered array of nanoholes of the substrate, and also on a glass substrate used as a reference. The base pressure before deposition was 5 x $10^{-7}$ Torr and the deposition pressure was kept at 3 mTorr using 20 sccm Ar flow and 50 Watts DC gun. Under this conditions the deposition rate was 3.5 nm/min. In order to avoid the cobalt oxidation a 2 nm Ta layer was deposited over the film.

### 2.2. *Structural and magnetic characterization*

The morphology of the samples was performed by a Scanning Electronic Microscope (SEM) Carl Zeiss EVO MA 10. The magnetic properties were measured by alternating-gradient force magnetometer (AGFM) at room temperature. Magnetization loops were measured with the applied field parallel to the antidots film plane.

### 2.3. *Micromagnetic Simulations*

Micromagnetic simulations were performed using the 3D OOMMF package [22]. Under this frame, the ferromagnetic system is divided into cubic cells with a uniform magnetization inside each cell. For our simulations we use the typical Cobalt parameters: saturation magnetization $M_S$= 1.4 x $10^6$ A/m, exchange stiffness constant A= 30 x $10^{-12}$ J/m, an anisotropy constant K = 0 $Jm^{-3}$ (since the film is polycrystalline) and a mesh size of 4 x 4 x 4 $nm^3$, where spins are free to rotate in three dimensions. In all the cases the damping constant is 0.5. In order to perform the micromagnetic simulations we have used two kinds of masks, the first one (from now on called the real mask) is obtained by exactly mimic the geometry of each antidot system. To achieve this, the SEM images (shown in figure 1 and corresponding to different pore diameters) have been treated by an image editor in order to improve the contrast and then they have been used as inputs in the simulation code [22]. The second kind of mask, from now on referred as the ideal mask, consist of a geometry that contains actual values for the pore diameter and inter-pore distance and a perfect hexagonal arrangement of the pores. In all the simulations we have used images of 1000 nm x

1000 nm and a film thickness of 28 nm. For the ideal mask a 50 nm interpore distance for 20 nm pores diameter and 100 nm interpore distance for 40 and 60 nm pores diameters were considered.

2.4. *First Order Reversal curves*

In order to study the magnetic properties of the samples in detail we applied the first order reversal curves (FORCs) technique [23]. Due to the fact that a hysteresis curve is a global measurement of the magnetization it does not clarify all the aspects involved in the magnetic properties of this system. In this context, the FORCs diagram technique makes an important contribution to the experimental study of magnetic systems composed of many magnetic particles or magnetic domains. In these cases FORCs diagrams provide information about the coercivity distribution and interactions inside the system and the portions of reversible and irreversible magnetization [24, 25]. This technique has been used to study diverse magnetic phenomena and systems [26, 27, 28, 29, 30, 31, 32].

Measuring a FORC begins by saturating the sample at high field. The field is then reduced to a *Ha* reversal field. Thus, a FORC is defined as the magnetization curve measured when the applied field increases from *Ha* until saturation. Magnetization for the *Hb* field over the FORC with the initial *Ha* field is denoted by *M* (*Ha*, *Hb*), where *Hb* > *Ha*. A statistical model that describes the system as a set of magnetic entities based on the Preisach model [33] is used to illustrate and obtain information from the FORCs. The probability density function of the ensemble is defined by the equation [34]:

$$\rho(H_a, H_b) \equiv -\frac{1}{2} \frac{\partial^2 M(H_a, H_b)}{\partial H_a \partial H_b} \qquad (1)$$

This function extends over the entire (*Ha*, *Hb*) plane. A FORCs diagram is a contour plot of equation 1, and can be expressed in terms of the variables Hc = (Hb + Ha)/2 and Hi = (Hb - Ha)/2, which are the commutation (coercivity of an entity) and interaction fields (shift of an entity) [33, 34], allowing us to capture the reversible magnetization component, which appears to be centered in Hc = 0. The density function in equation 1 for a sample is obtained by numeric derivation of the *M* (*Ha*, *Hb*) function, which contains all the measured FORCs. In this way, the FORCs diagram is obtained by making a contour plot of the equation 1.

## 3. Results and discussions

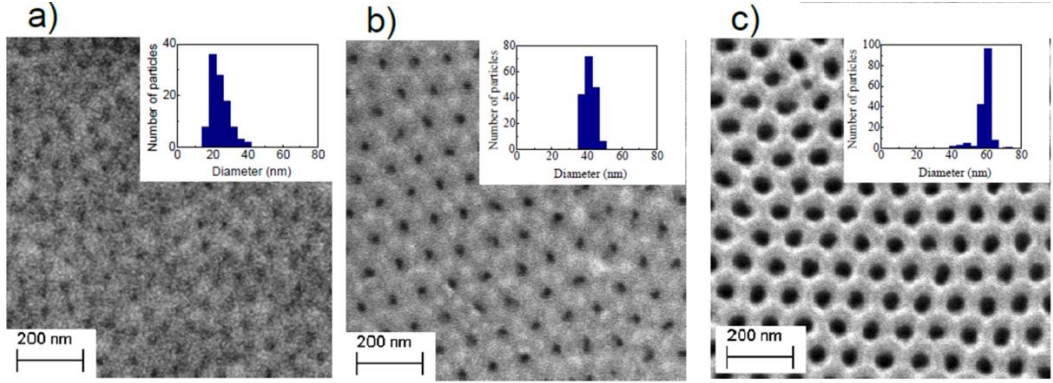

**Figure 1**. Scanning Electron Microscope images of cobalt antidots where the pore diameters *d* are (a) 20 nm, (b) 40 nm and (c) 60 nm. The inset graph shows the diameter size distribution for each sample.

The morphology of the arrays of cobalt antidots, obtained with the process described in section 2.1, are shown in figure 1. It can be seen that the sputtered cobalt film adopts the topology imposed by the substrates (i.e. the anodized alumina membranes). In this way, non-magnetic inclusions, characterized by the membrane pore diameter, arise and make the film a patterned medium. The inset graph in figure 1 shows the diameter dispersion of the antidots samples, here is possible to see the good uniformity and distribution of the characteristic geometrical parameters (pore diameter, interpore distance and hexagonal hole arrangement) due to high quality anodization process.

As we detailed in section 2.3, micromagnetic simulation have been performed by both, taking the SEM images shown in figure 1 - using them as masks – in one hand, and making a defect-free perfect image of the hexagonal arrangements (keeping the pore diameter and interpore distance), in the other hand, with the aim to decorrelate the pore diameter effect from hexagonal ordering. Figure 2 shows the remanent and coercivity states for both real and ideal mask obtained of such simulation.

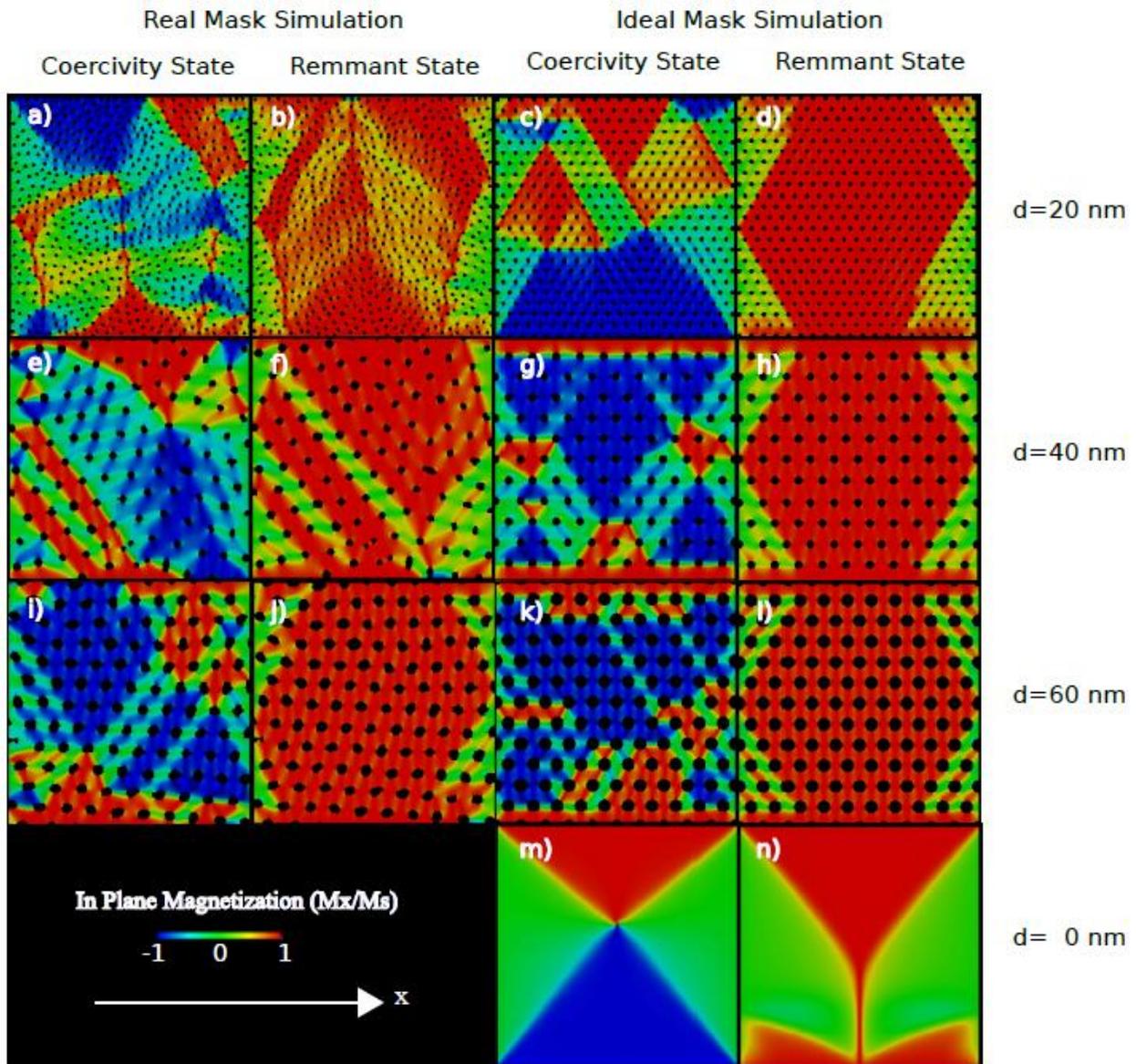

**Figure 2**. Micromagnetic simulation snapshots for the real (first and second columns) and the ideal (third and fourth columns) masks for the different pore diameters *d*. The fourth row corresponds to the continuous film. The colour code corresponds to the magnetization component along the applied field.

The first and second columns in figure 2 show six particular snapshots, corresponding to the real mask simulations. In these images, the colour code corresponds to the magnetization component along the applied field (the *x* direction in this case). The images shown in the second column, (b), (f), and (j), are the magnetic configurations in the remnant magnetization state ($H = 0$). According to the colour code, green zones have no magnetization component along the applied field. In this way, according to the simulations, the first stages of the magnetization reversal take place at the edges of the structures. Besides, for the structures having pore diameters of 40 and 20 nm the magnetization starts rotating around the pores located at the centre of the structure. This seems to indicate that the relative preponderance of the main reversal mechanisms (i.e. coherent rotation and domain wall propagation) is affected by the pore diameter; being the domain wall propagation dominant in the largest pore diameter.

Another feature than can be seen from the remnant state simulations is that, for the three pore diameters, most of the magnetization is still aligned to the initial saturation direction. The reduced (or normalized) remanent magnetization $m_R$ ($m_R = M_R/M_S$) is found to be 0.85, 0.72 and 0.75 for 60, 40 and 20 nm pore diameter respectively. These values have been extracted from the field dependent magnetization loops, obtained in the same micromagnetic simulation, shown in figure 3.

Let's consider now the magnetic configurations at the coercive field value, shown in figure 2. From snapshot (i), corresponding to a 60 nm pore diameter, it can be seen that the adopted configuration exhibits a well-defined regularity where the domains shape display a rhomboid nature. Such structure is consistent with the presence of a six-folded anisotropy. Wang et al showed that a six-fold shape anisotropy arises in permalloy antidot structures where the pores are arranged in a honeycomb structure [35]. In these hexagonally ordered pore geometries, each of the directions linking one defined pore to its first neighbors corresponds to an easy anisotropy axis, whereas the directions linking the pore to its second neighbors will be associated to a hard anisotropy axis. Since green stripes in figure 2(i) are regions with no magnetization component along the applied field, we can conclude that in such zones the magnetization points either in the + y direction or the - y direction, this is, along the easy axis which turns out to be perpendicular to the applied field direction. On the other hand, yellow stripes, which mostly go from a pore to its first neighbor, being therefore lying along hard directions, are associated to the presence of domain walls. In the case of figure 2(e) (40 nm pore diameter) fewer domain walls are present and the size of domains is larger. These domains seem to be oriented rather locally since the hexagonal order is partially lost due to the presence of defects. For the smallest pore diameter, shown in figure 2(a), almost no domain walls are present between neighboring pores; instead, 90° domain walls separate large domains oriented parallel to the borders of the overall structure.

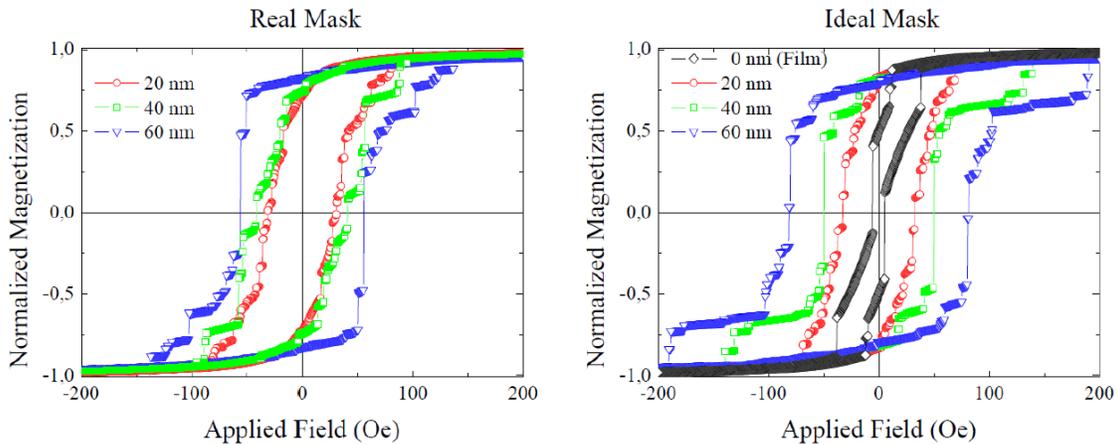

**Figure 3**. Field dependent magnetization loops of cobalt antidots obtained via micromagnetic simulations using the real masks (left) and ideal mask (right).

Figure 2 also shows the magnetic configurations found at the remnant state (fourth column) as well as at the coercive field (third column) for ideal mask. Together with the results obtained the magnetic configurations for a 28 nm thick continuous film are also presented. From these images it can be seen that, even if the hole density is the same (100 holes / $\mu m^2$) for the different pore diameters, a bigger pore diameter is associated to the presence of more domains and thus to a higher density of domain walls. Owing to the hexagonal arrangement, six first neighbors and six second neighbors surround each hole. In this way, every hole configures six constrictions with its first neighbors. Domain walls can be trapped or pinned by these constrictions. Each of these domain walls will experience a spatially variant dipolar interaction along the whole structure and thus it will be subjected to a landscape of pinning potentials [36]. A domain wall trapped in a pinning potential can adopt several stable configurations and can become unpinned through a complex process where a depinning field distribution arises [37, 38].

From third column at figure 2 it can be seen that as the pore diameter increase, a higher domain wall density was observed. That can be interpreted as a decreasing in the pinning potential experienced by domain wall. This should be reflected in a drop of the coercive field as the pore diameter increases. Such a drop is found from the M(H) curves obtained via micromagnetic simulations and showed in figure 3 and the same tendency was found in the simulated hysteresis curves obtained with the real masks. However, the values for the coercive field obtained with the real geometry are lower than those obtained with the ideal geometry (see figure 4). This difference can arise from two distinct features in which the ideal and real geometry differ from each other: the arrangement and the exact geometry of the hole. In a real arrangement defects appear resulting in a breaking of the hexagonal symmetry, changing thus the number of neighbors, typically lowering this value from 6 to 5, and changing therefore the number of domain walls trapped around a hole. Another effect of the symmetry breaking is that the six-fold long range shape anisotropy is lost, leading to less defined domains as can be seen in figure 2. On the other hand, the exact geometry of the hole is crucial since it determines the shape of the constriction and therefore the punctual pinning potential experienced by the domain walls, playing thus a major role in the coercive field [39, 40, 41, 42]. It is worth to note that the sample with pores of 20 nm has an inter-pore distance of 50 nm, in comparison to 100 nm in the other two samples. However, as the pore diameter increase, the behavior of $H_C$ and Mr change. Simulations made from an ideal mask have bigger coercive value than the simulations made from a real mask, and in fig. 4 the real mask have a behavior most similar to the experiment. One of the principal reasons for the differences between the experimental data and the simulated data is the fact that the simulation is performed at 0 K, but the constants have their values at room temperature; moreover the size of the samples simulated are smaller compared with the experimental measurements.

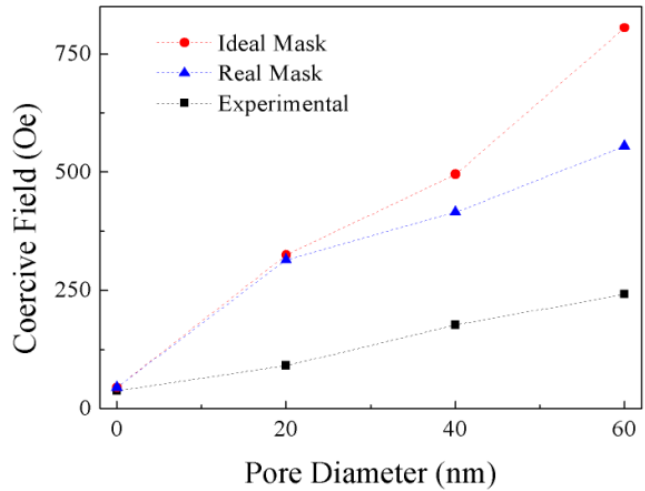

**Figure 4**. Coercivity values obtained from both real and ideal simulations and from experiments, corresponding to the different pore diameters.

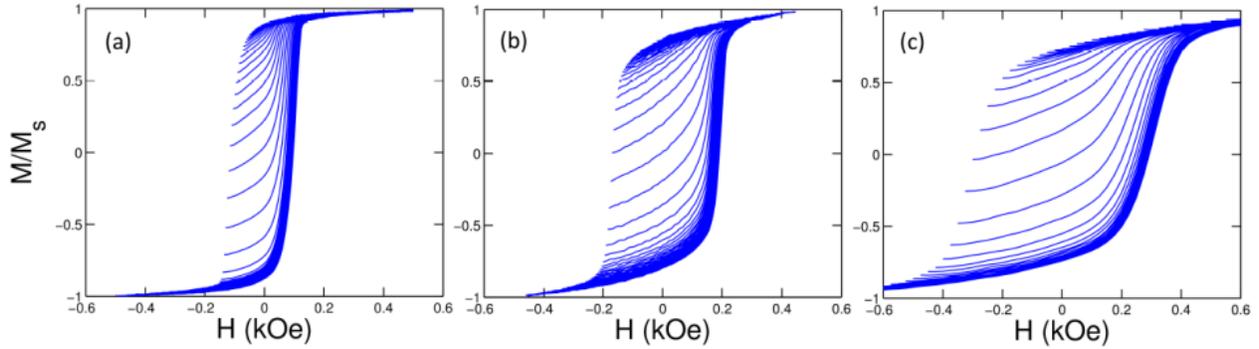

**Figure 5**. First-order reversal curves corresponding to an antidot array with a a) 20 nm , b) 40 nm and a) 60 nm pore diameter.

In order to get a deeper insight on the magnetization reversal and the different possible depinning mechanisms in the experimental samples, first order reversal curves have been measured and the diagrams are analyzed. Figure 5 shows the FORCs measured with the external magnetic field applied parallel to the plane of the entire sample set. The contour delineated by the FORCs corresponds exactly to the hysteresis curve. In order to evidence the differences between the systems, it is necessary to carry out FORCs diagrams, as shown in figure 6.

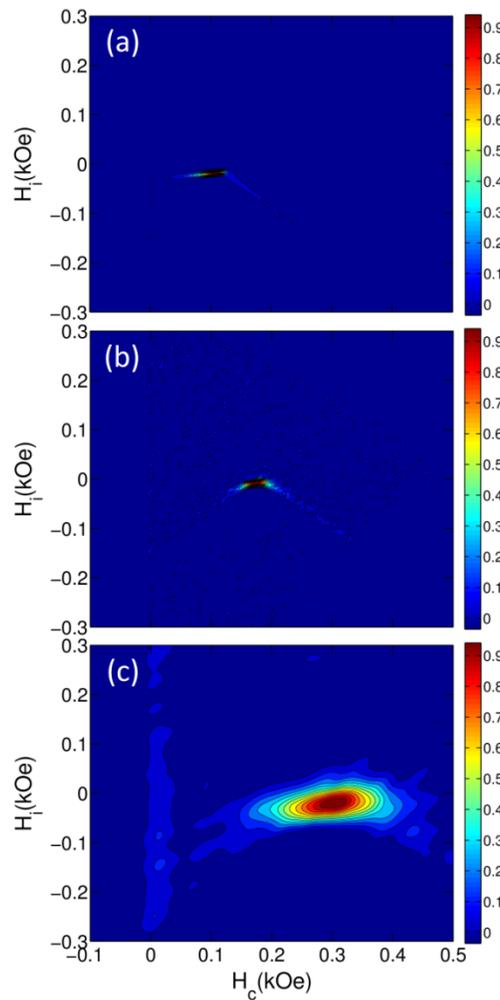

**Figure 6**. First-order reversal diagrams corresponding to pore diameters of: (a) 20nm, (b) 40nm and (c) 60nm.

The FORCs diagram for the thin film with 20 nm pore diameter shows a small spot of irreversible magnetization (for Hc > 0 Oe), with a narrow distribution of interaction and a small distribution of coercivity centered in Hc = 106 Oe, a value close to that of the hysteresis curve of 91 Oe. The small spot on the diagram indicates that the system has a more homogeneous structure of magnetic domains than the system with larger pores, as the simulation shows (see figure 1). Theoretically, the FORCs diagram of a bi-stable monodomain is a point located over the coercivity axis in a value equal to the coercivity of the system.

The diagram of the system with 40 nm pores shows a similar panorama to that of the 20 nm pore system: only a spot of irreversible magnetization centered in Hc = 176 Oe, a value very close to the hysteresis curve of 178 Oe. This spot shows slightly higher interaction and coercivity distributions than those of the 20 nm pore system. Specifically, its shows an increase in the width of the interaction distribution, which indicates the presence of a more complex magnetic structure than that of the 20 nm pore system and a larger number of magnetic domains interacting. This diagram presents noise because the magnetic signal of this sample was captured with more noise than the 20 and 60 nm samples.

If we take a look now to the diagram corresponding to the 60 nm pore, shown in figure 6(c), we can see that a large irreversible magnetization spot centered in Hc=300 Oe is obtained. This system has larger coercivity and interaction distributions than the other two systems (20 and 40 nm), which indicates the presence of more complex magnetic structure than those of the systems of smaller pores. The increase in coercivity distribution indicates that there are different magnetic regions that revert their magnetization at distinct external fields, which results in a denser and more complex domain structure, with smaller magnetic domains, and a labyrinth structure of domains as shown in the simulation (see figure 2). In turn, the increased number of magnetic regions oriented in different directions increases magnetic interaction among neighboring regions, which is evidenced in the increased interaction distribution in the diagram. Finally, in all the diagrams (for 20, 40, and 60 nm) we observe considerably greater distribution of coercivity than of interaction, which indicates the predominant effect of pores with larger diameters is the formation of a larger number of magnetic domains. Given the two-dimensional nature of the samples and the direction of the external magnetic field, the effect of magnetic interaction is lower than that of the formation of magnetic domains.

## 4. Conclusions

Arrays of Co antidots with different pore diameters have been fabricated and the magnetic properties were analyzed experimentally and by means of numerical simulations. Simulations of real and ideal masks were used to investigate the effect of the pore sizes and how it affects the magnetic properties. The analysis of the FORCs diagrams shows an increase in the coercivity and interaction field distributions in the samples with larger pores. From the micromagnetic simulations and magnetic characterizations we can conclude that as the size of the pores increase, and the space for the propagation of the domain walls is reduced, interaction among different propagating domains produces multiple smaller domains and consequently gives rise to a more complex magnetic domain structures, which results in an increase in the coercivity of the films.

## 5. Acknowledgments

Support from FONDECYT under projects 3120059, 1110252, 1110784, 11110130, 3130397 and 3130393; USAFOSR Award No. FA9550-11-1-0347; Grant ICM P10-061-F by Fondo de Innovacion para la Competitividad-MINECON; and Financiamiento Basal para Centros Científicos y Tecnológicos de Excelencia under project FB0807 are gratefully acknowledged.